\documentclass[12pt,letterpaper]{article}
\usepackage{epsfig,rotating,setspace,latexsym,amsmath,epsf,amssymb,amsfonts,bm,theorem,cite,subfigure}
\usepackage{algorithm,algorithmic,graphicx,epsf,authblk,epstopdf,url,color,multirow}
\newtheorem{theorem}{Theorem}

\newtheorem{lemma}{Lemma}
\newenvironment{Proof}[1]{\medskip\par\noindent{\bf Proof:\,}\,#1}{{\mbox{\,$\blacksquare$}\par}}

\newcommand{\bx}{{\mathbf{x}}}
\newcommand{\bw}{{\mathbf{w}}}
\newcommand{\by}{{\mathbf{y}}}
\newcommand{\bh}{{\mathbf{h}}}
\newcommand{\cq}{{\mathcal{Q}}}

\begin{document}
	
	\title{The Capacity of Private Information Retrieval from Coded Databases\thanks{This work was supported by NSF Grants CNS 13-14733, CCF 14-22111, CCF 14-22129, and CNS 15-26608.}}
	
	\author{Karim Banawan \qquad Sennur Ulukus\\
		\normalsize Department of Electrical and Computer Engineering\\
		\normalsize University of Maryland, College Park, MD 20742 \\
		\normalsize {\it kbanawan@umd.edu} \qquad {\it ulukus@umd.edu}}
	
	\maketitle
	
\begin{abstract}
		
	We consider the problem of private information retrieval (PIR) over a distributed storage system. The storage system consists of $N$ non-colluding databases, each storing a coded version of $M$ messages. In the PIR problem, the user wishes to retrieve one of the available messages without revealing the message identity to any individual database. We derive the information-theoretic capacity of this problem, which is defined as the maximum number of bits of the desired message that can be privately retrieved per one bit of downloaded information. We show that the PIR capacity in this case is $C=\left(1+\frac{K}{N}+\frac{K^2}{N^2}+\cdots+\frac{K^{M-1}}{N^{M-1}}\right)^{-1}=(1+R_c+R_c^2+\cdots+R_c^{M-1})^{-1}=\frac{1-R_c}{1-R_c^M}$, where $R_c$ is the rate of the $(N,K)$ code used. The capacity is a function of the code rate and the number of messages only regardless of the explicit structure of the storage code. The result implies a fundamental tradeoff between the optimal retrieval cost and the storage cost. The result generalizes the achievability and converse results  for the classical PIR with replicating databases to the case of coded databases.     
		
\end{abstract}
	
\section{Introduction}

Protecting the privacy of downloaded information from curious publicly accessible databases has been the focus of considerable research within the computer science community \cite{ChorPIR, PIRsurvey2004 ,ostrovsky2007survey,yekhanin2010private}. Practical examples for this problem include: ensuring privacy of investors upon downloading records in a stock market, and ensuring the privacy of activists against authoritarian regimes while browsing restricted contents from the internet, see \cite{RazanPIR,ChorPIR}.  In the seminal paper Chor et. al. \cite{ChorPIR}, the classical problem of private information retrieval (PIR) is introduced. In the classical PIR setting, a user requests to download a certain message (or file) from $N$ non-communicating databases without leaking the identity of the message to any individual database. The contents of these databases are identical, i.e., they are repetition coded. A trivial solution for this seemingly challenging task is to download all of the contents of the databases. However, this solution is highly impractical, in particular for large number of messages which is the case in modern storage systems. The aim of the PIR problem is to design efficient retrieval schemes that maximize the ratio of the desired information bits to the total downloaded bits under the privacy constraint.  

In the classical PIR problem, the user prepares $N$ queries each directed to a specific database. The queries are designed such that they do not reveal any information about the identity of the desired message. Upon receiving these queries, databases respond truthfully with answering strings. Based on the collected answer strings, the user reconstructs the desired message. In the original formulation of the problem in the computer science literature \cite{ChorPIR}, the messages are assumed to have a size of one bit. In this formulation, the performance metric was the sum of lengths of the answer strings (download cost) and the size of the queries (upload cost). The information-theoretic reformulation of the problem assumes that the messages are of arbitrarily large size and hence the upload cost can be neglected with respect to the download cost \cite{YamamotoPIR}. The pioneering work \cite{JafarPIR} derives the exact capacity of the classical PIR problem. The capacity is defined as the maximum number of bits of the desired message per bit of total download. The achievable scheme is based on an interesting relationship between PIR and blind interference alignment introduced for wireless networks in \cite{BIA} as observed in \cite{JafarPIRBlind}. \cite{JafarColluding} extends this setting to the case of $T$ colluding databases with and without node failures. The main difference from the non-colluding case is that the user asks for MDS-coded versions of the contents of the databases. Another interesting extension of the problem is the symmetric PIR \cite{symmetricPIR}, in which the privacy of the undesired messages need to be preserved against the user. 

Due to node failures and erasures that arise naturally in any storage system, redundancy should be introduced \cite{StorageSurvey}. The simplest form of redundancy is repetition coding. Although repetition coding across databases offers the highest immunity against erasures and the simplicity in designing PIR schemes, it results in extremely large storage cost. This motivates the use of erasure coding techniques that achieve the same level of reliability with less storage cost. A common erasure coding technique is the MDS code that achieves the optimal redundancy-reliability tradeoff. An $(N,K)$ MDS code maps $K$ sub-packets of data into $N$ sub-packets of coded data. This code tolerates upto $N-K$ node failures (or erasures). By connecting to any $K$ storage nodes, the node failure can be repaired. Despite the ubiquity of work on the classical PIR problem, little research exists for the coded PIR to the best of our knowledge with a few exceptions: \cite{RamchandranPIR} which has initiated the work on coded databases and has designed an explicit erasure code and PIR algorithm that requires only one extra bit of download to provide perfect privacy. The result is achieved in the expense of having storage nodes that grow with the message size. \cite{YamamotoPIR} considers a general formulation for the coded PIR problem, and obtains a tradeoff between storage and retrieval costs based on certain sufficient conditions. \cite{RazanPIR} presents the best known achievable scheme for the MDS-coded PIR problem, which achieves a retrieval rate of $R = 1-R_c$, where $R_c$ is the code rate of the storage system. The scheme is universal in that it depends only on the code rate. Finally, \cite{VardyPIR} investigates the problem from the storage overhead perspective and shows that information-theoretic PIR can be achieved with storage overhead arbitrarily close to the optimal value of 1 by proposing new binary linear codes called the $k$-server PIR codes.

In this paper, we consider the PIR problem for non-colluding and coded databases. We use the information-theoretic formulation. We do not assume any specific structure on the generator matrix of the distributed storage code other than linear independence of every $K$ columns. This condition is equivalent to restricting the storage code structure to MDS codes. Note also that the dimensions of the generator matrix $(N,K)$ are not design parameters that can grow with the message size as in \cite{RamchandranPIR}. This formulation includes the models of \cite{JafarPIR} and \cite{RazanPIR} as special cases. We show that the exact PIR capacity in this case is given by $C=\left(1+\frac{K}{N}+\frac{K^2}{N^2}+\cdots+\frac{K^{M-1}}{N^{M-1}}\right)^{-1}=(1+R_c+R_c^2+\cdots+R_c^{M-1})^{-1}=\frac{1-R_c}{1-R_c^M}$. The PIR capacity depends only on the code rate $R_c$ and the number of messages $M$ irrespective of the generator matrix structure or the number of nodes. Surprisingly, the result implies the optimality of separation between the design of the PIR scheme and the storage code for a fixed code rate. The result outperforms the best-known lower bound in \cite{RazanPIR}. The result reduces to the repetition-coded case in \cite{JafarPIR} by observing that $R_c=\frac{1}{N}$ in that case. The achievable scheme is similar to the scheme in \cite{JafarPIR} with extra steps that entail decoding of the interference and the desired message by solving $K$ linearly independent equations. The converse proof hinges on the fact that the contents of any $K$ storage nodes are independent and hence the answer strings in turn are independent. We prove the base induction step, i.e., the case $M=2$, in a more direct way than \cite{JafarPIR} and generalize the inductive relation in \cite{JafarPIR} to account for coding. We  present two new lemmas that capture the essence of the converse proof, namely: interference lower bound for $M=2$, and interference conditioning for general $M$.

\allowdisplaybreaks

\section{System Model}
\begin{figure}[t]
	\centering
	\includegraphics[width=0.6\textwidth]{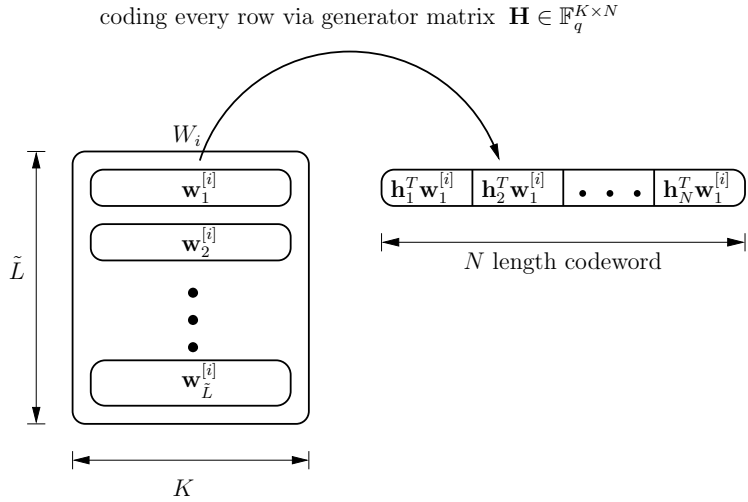}
	\caption{Coding process for message $W_i$.}
	\label{storage_code}
	\vspace*{-0.4cm}
\end{figure}

\begin{figure}[t]
	\centering
	\includegraphics[width=0.9\textwidth]{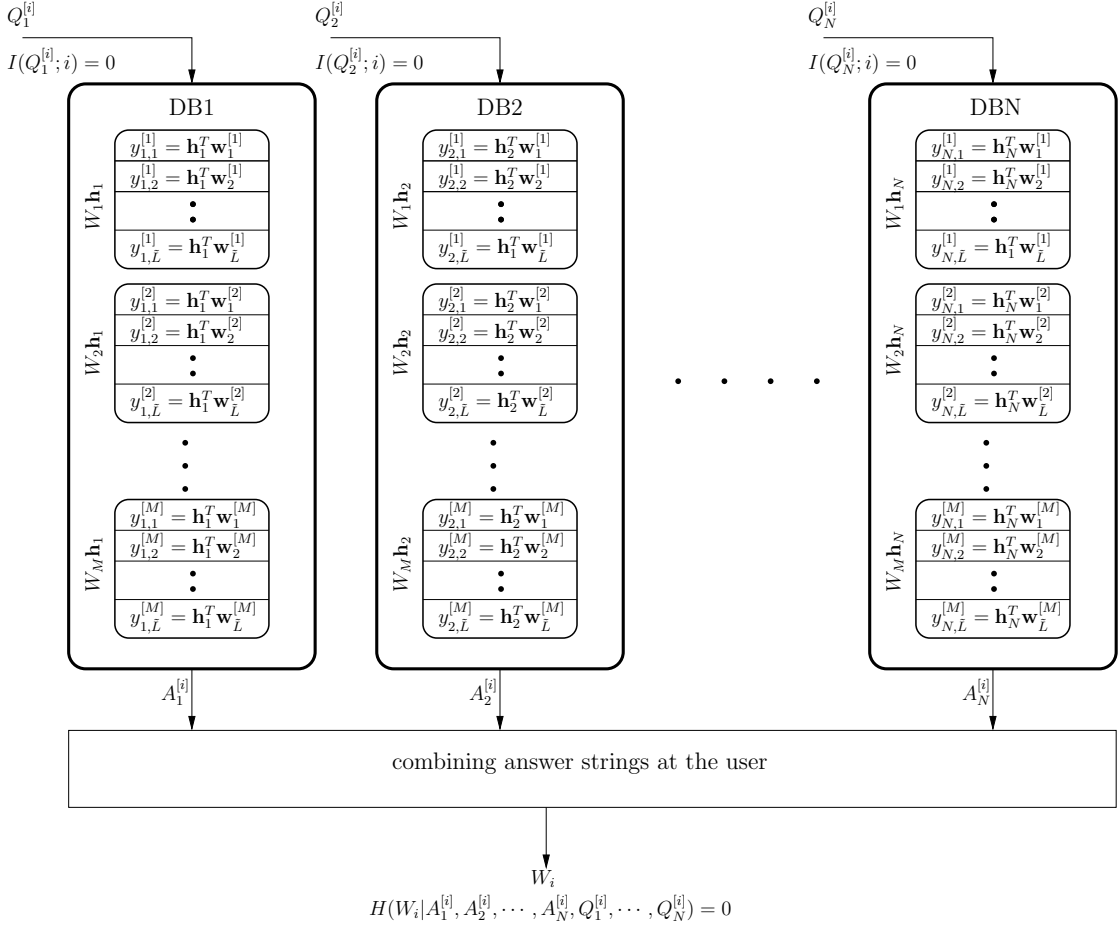}
	\caption{System model of the coded PIR problem.}
	\label{PIR_model}
	\vspace*{-0.4cm}
\end{figure}

Consider a linear $(N,K)$ distributed storage system storing $M$ messages (or files). The messages are independent and identically distributed with
\begin{align}
H(W_i)&=L, \quad  i \in \{1, \cdots, M\} \\
H(W_1,W_2,\cdots, W_M)&=ML
\end{align}
The message $W_i, \: i \in \{1, \cdots, M\}$ is a $\mathbb{F}_q^{\tilde{L} \times K}$ matrix with sufficiently large field $\mathbb{F}_q$, such that $\tilde{L} \times K=L$. The elements of $W_i$ are picked uniformly and independently from $\mathbb{F}_q$. We denote the $j$th row of message $W_i$ by $\bw_j^{[i]} \in \mathbb{F}_q^{K}$. The generator matrix of the $(N,K)$ storage code $\mathbf{H}$ is a $\mathbb{F}_q^{K \times N}$ matrix such that
\begin{align}
\mathbf{H}=
\begin{bmatrix}
\bh_1 & \bh_2 & \cdots & \bh_N
\end{bmatrix}_{K \times N}
\end{align}
where $\bh_i \in \mathbb{F}_q^K, \: i \in \{1, \cdots, N\}$. In order to have a feasible storage code, we assume that any set $\mathcal{K}$ of columns of $\mathbf{H}$ such that $|\mathcal{K}| \leq K$ are linearly independent. The storage code $f_n: \bw_j^{[i]} \rightarrow y_{n,j}^{[i]}$ on the $n$th database maps each row of $W_i$ separately into coded bit  $y_{n,j}^{[i]}$, see Fig.~\ref{storage_code},
\begin{align}
y_{n,j}^{[i]}=\bh_n^T \bw_j^{[i]}
\end{align}
Consequently, the stored bits $\by_n \in \mathbb{F}_q^{M\tilde{L}}$ on the $n$th database, $n \in \{1,\cdots, N\}$ are concatenated projections of all messages $\{W_1,\cdots, W_M\}$ and are given by
\begin{align}
\by_n&=\begin{bmatrix}
W_1 \\
\vdots\\
W_M
\end{bmatrix} \bh_n \\
&=\begin{bmatrix}
\bh_n^T \bw_1^{[1]} &
\dots &
\bh_n^T \bw_{\tilde{L}}^{[1]}&
\bh_n^T \bw_1^{[2]}	 &
\dots &
\bh_n^T \bw_{\tilde{L}}^{[2]}&
\dots &
\bh_n^T \bw_1^{[M]}	 &
\dots &
\bh_n^T \bw_{\tilde{L}}^{[M]}
\end{bmatrix}^T
\end{align}
The explicit structure of the coded storage system is illustrated in Table~\ref{table:code}.
\begin{table}[]
	\centering
	\caption{Explicit structure of $(N,K)$ code for distributed databases with $M$ messages.}
	\label{table:code}
	\begin{tabular}{|l|c|c|c|c|}
		\hline
		& DB1 ($\by_1$)                   & DB2 ($\by_2$)                   &\hspace{0.4cm} $\cdots$ \hspace{0.4cm} & DBN ($\by_N$)                   \\ \hline
		\rule{0pt}{3ex}  
		\multirow{4}{*}{\rotatebox[origin=c]{90}{\parbox[c]{2cm}{\centering message 1}}}   & $\bh_1^T \bw_1^{[1]}$           & $\bh_2^T \bw_1^{[1]}$           & \hspace{0.5cm} $\cdots$ \hspace{0.4cm} & $\bh_N^T \bw_1^{[1]}$           \\
		& $\bh_1^T \bw_2^{[1]}$           & $\bh_2^T \bw_2^{[1]}$           &\hspace{0.4cm} $\cdots$ \hspace{0.4cm} & $\bh_N^T \bw_2^{[1]}$           \\
		& $\vdots$                        & $\vdots$                        &\hspace{0.4cm} $\cdots$ \hspace{0.4cm} & $\vdots$                        \\
		& $\bh_1^T \bw_{\tilde{L}}^{[1]}$ & $\bh_2^T \bw_{\tilde{L}}^{[1]}$ & \hspace{0.4cm} $\cdots$ \hspace{0.4cm} & $\bh_N^T \bw_{\tilde{L}}^{[1]}$    \rule[-1.2ex]{0pt}{0pt}\\ \hline
		\rule{0pt}{3ex}  
		\multirow{4}{*}{\rotatebox[origin=c]{90}{\parbox[c]{2cm}{\centering message 2}}}   & $\bh_1^T \bw_1^{[2]}$           & $\bh_2^T \bw_1^{[2]}$           & \hspace{0.5cm} $\cdots$ \hspace{0.4cm} & $\bh_N^T \bw_1^{[2]}$           \\
		& $\bh_1^T \bw_2^{[2]}$           & $\bh_2^T \bw_2^{[2]}$           & \hspace{0.4cm} $\cdots$ \hspace{0.4cm} & $\bh_N^T \bw_2^{[2]}$           \\
		& $\vdots$                        & $\vdots$                        & \hspace{0.4cm} $\cdots$ \hspace{0.4cm} & $\vdots$                        \\
		& $\bh_1^T \bw_{\tilde{L}}^{[2]}$ & $\bh_2^T \bw_{\tilde{L}}^{[2]}$ &\hspace{0.4cm} $\cdots$ \hspace{0.4cm} & $\bh_N^T \bw_{\tilde{L}}^{[2]}$    \rule[-1.2ex]{0pt}{0pt}\\ \hline
		\rule{0pt}{5ex}
		$\vdots$                     & $\vdots$                        & $\vdots$                        & \hspace{0.5cm} $\cdots$ \hspace{0.5cm} & $\vdots$                       \rule[-3ex]{0pt}{0pt} \\ \hline
		\rule{0pt}{3ex}
		\multirow{4}{*}{\rotatebox[origin=c]{90}{\parbox[c]{2.5cm}{\centering message $M$}}} & $\bh_1^T \bw_1^{[M]}$           & $\bh_2^T \bw_1^{[M]}$           &\hspace{0.5cm} $\cdots$ \hspace{0.5cm} & $\bh_N^T \bw_1^{[M]}$           \\
		& $\bh_1^T \bw_2^{[M]}$           & $\bh_2^T \bw_2^{[M]}$           & \hspace{0.5cm} $\cdots$ \hspace{0.5cm} & $\bh_N^T \bw_2^{[M]}$           \\
		& $\vdots$                        & $\vdots$                        & \hspace{0.5cm} $\cdots$ \hspace{0.5cm} & $\vdots$                        \\
		& $\bh_1^T \bw_{\tilde{L}}^{[M]}$ & $\bh_2^T \bw_{\tilde{L}}^{[M]}$ & \hspace{0.5cm} $\cdots$ \hspace{0.5cm} & $\bh_N^T \bw_{\tilde{L}}^{[M]}$ \rule[-1.2ex]{0pt}{0pt}\\ \hline
	\end{tabular}
\end{table}
The described storage code can tolerate up to $N-K$ errors by connecting to any $K$ databases. Thus, we have for any set $\mathcal{K}$ such that $|\mathcal{K}| \geq K$,
\begin{align}\label{decodeK}
H(\by_{\bar{\mathcal{K}}}|\by_{\mathcal{K}})=0
\end{align}
where $\by_{\mathcal{K}}$ are the stored bits on databases indexed by $\mathcal{K}$, and $\bar{\mathcal{K}}$ is the complement of the set $\mathcal{K}$. The code rate of this distributed storage system $R_c$ is given by
\begin{align}
R_c=\frac{K}{N}
\end{align}
The retrieval process over coded databases is illustrated in Fig.~\ref{PIR_model}. To retrieve $W_i$, the user generates a query $Q_n^{[i]}$ and sends it to the $n$th database. Since the user does not have knowledge about the messages in advance, the queries are independent of the messages, 
\begin{align}
I(Q_1^{[i]},\cdots, Q_N^{[i]};W_1, \cdots, W_M)=0
\end{align}
In order to ensure privacy, the queries should be independent of the desired message index $i$, i.e., the privacy constraint is,
\begin{align}
I(Q_n^{[i]};i)=0, \quad  n \in \{1, \cdots, N\}
\end{align}
Each database responds with an answer string $A_n^{[i]}$, which is a deterministic function of the received query and the stored coded bits in the $n$th database. Hence, by the data processing inequality,
\begin{align}
H(A_n^{[i]}|Q_n^{[i]},\by_n)=H(A_n^{[i]}|Q_n^{[i]},W_1,\cdots, W_M)=0
\end{align}
In addition, the user should be able to decode $W_i$ reliably from all the answer strings collected from the $N$ databases. Consequently, we have the following reliability constraint,
\begin{align}\label{reliability}
H(W_i|A_1^{[i]},\cdots,A_N^{[i]},Q_1^{[i]},\cdots,Q_N^{[i]})=0
\end{align}
The retrieval rate $R$ for the PIR problem is the ratio of the size of the desired message to the total download cost,
\begin{align}\label{PIRrate}
R=\frac{H(W_i)}{\sum_{n=1}^N H(A_n^{[i]})}
\end{align}
The PIR capacity $C$ is the supremum of $R$ over all retrieval schemes. 

In this paper, as in \cite{JafarPIR}, we follow a Shannon theoretic formulation by assuming that the message size can be arbitrarily large. Also, we neglect the upload cost with respect to the download cost as in \cite{JafarPIR}.

We note that the described storage code is a generalization of the repetition-coded problem in \cite{JafarPIR}. If $K=1$ and $h_n=1,\:  n \in \{1, \cdots, N\}$, then the problem reduces to the classical PIR in \cite{JafarPIR}. In addition, the systematic MDS-coded problem in \cite{RazanPIR} is a special case of this setting with $\bh_n=\mathbf{e}_n, \: n \in {1, \cdots, K}$, where $\mathbf{e}_n$ is the $n$th standard basis vector.
\section{Main Result}
\begin{theorem}\label{thm1}
	For an $(N,K)$ coded distributed database system with coding rate $R_c=\frac{K}{N}$ and $M$ messages, the PIR capacity is given by
	\begin{align}\label{PIR_rate}
	C &= \frac{1-R_c}{1-R_c^M} \\
	  &= \frac{1}{1+R_c+\cdots+R_c^{M-1}} \\
	  &= \left(1+\frac{K}{N}+\frac{K^2}{N^2}+\cdots+\frac{K^{M-1}}{N^{M-1}}\right)^{-1}
	\end{align}
\end{theorem}

\begin{figure}[t]
	\centering
	\includegraphics[width=0.7\textwidth]{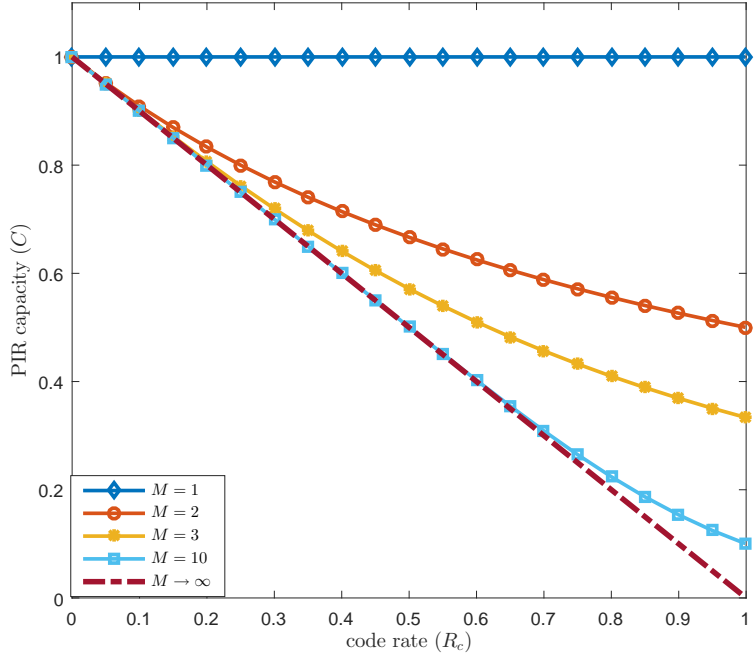}
	\caption{PIR capacity versus $R_c$.}
	\label{fig:capacity}
	\vspace*{-0.4cm}
\end{figure}

We have the following remarks about the main result. We first note that the PIR capacity in (\ref{PIR_rate}) is a function of the coding rate $R_c$ and the number of messages $M$ only, and does not depend on the explicit structure of the coding scheme (i.e., the generator matrix) or the number of databases. This observation implies the universality of the scheme over any coded database system with the same coding rate and number of messages. The result also entails the optimality of separation between distributed storage code design and PIR scheme design for a fixed $R_c$. We also note that the capacity $C$ decreases as $R_c$ increases. As $R_c \rightarrow 0$, the PIR capacity approaches $C=1$. On the other hand, as $R_c \rightarrow 1$, the PIR capacity approaches $\frac{1}{M}$ which is the trivial retrieval rate obtained by downloading the contents of all databases. This observation conforms with the result of \cite{YamamotoPIR}, in which a fundamental trade off exists between storage cost and the retrieval download cost. The capacity expression in Theorem~1 is plotted in Fig.~\ref{fig:capacity} as a function of the code rate $R_c$ for various numbers of messages $M$. 

The capacity in (\ref{PIR_rate}) is strictly larger than the best-known achievable rate in \cite{RazanPIR}, where $R=1-R_c$ for any finite number of messages. We observe also that the PIR capacity for a given fixed code rate $R_c$ is monotonically decreasing in $M$. The rate in (\ref{PIR_rate}) converges to $1-R_c$ as $M \rightarrow \infty$. Intuitively, as the number of messages increases, the undesired download rate must increase to hide the identity of the desired message; eventually, the user should download all messages as $M \rightarrow \infty$. Our capacity here generalizes the capacity in \cite{JafarPIR} where $R_c=\frac{1}{N}$. That is, the classical PIR problem may be viewed as a special case of the coded PIR problem with a specific code structure which is repetition coding. 
\section{Achievability Proof}
In this section, we present the general achievable scheme for Theorem~\ref{thm1}. We give a few specific examples in Section~\ref{sec:examples}. Our achievable scheme generalizes the achievable scheme in \cite{JafarPIR} which induces symmetry across databases and symmetry across messages, and exploits the side information. The achievable scheme here includes two extra steps due to the presence of coding: decoding of the interference and decoding of the desired rows which are not present in \cite{JafarPIR}.
\subsection{Achievable Scheme}
The scheme requires $\tilde{L}=N^M$, which implies that the size of message $H(W_i)=L=KN^M$. The scheme is completed in $M$ rounds, each corresponding to the sum of $i$ terms, $i \in \{1,\cdots, M\}$, and is repeated $K$ times to decode the desired message; see Tables~\ref{code(5,3)} and \ref{code(3,2)} for examples.
\begin{enumerate}
	\item \textit{Index preparation:} The user interleaves the indices of rows for all messages randomly and independently from each other, i.e., for any message $W_m$,
	\begin{equation}
	\bx_i^{[m]}=\bw_{\pi(i)}^{[m]}, \quad  i \in \{1, \cdots, \tilde{L}\}
	\end{equation}
	where $\pi(\cdot)$ is a random interleaver known privately to the user only. In this case the rows chosen at any database appear to be chosen at random and independent from the desired message index.
	\item \textit{Initialization:} The user downloads $K^{M-1}$ desired coded PIR bits from different rows from database 1 (DB1) and sets round index $i=1$.
	\item \textit{Symmetry across databases:} The user downloads  $K^{M-1}$ desired bits each from a different row from each database. Then, the total number of desired bits in the $i$th round is $NK^{M-1}$.
	\item \textit{Message symmetry:} To satisfy the privacy constraint, the user needs to download an equal amount of coded bits from all other messages. Consequently, the user downloads $\binom{M-1}{i}K^{M-i}(N-K)^{i-1}$ bits from each database. The undesired equation is a sum of $i$ terms picked from the remaining undesired messages. Hence, the number of undesired equations downloaded in the $i$th round is $N\binom{M-1}{i}K^{M-i}(N-K)^{i-1}$.
	\item \textit{Decoding the interference:} The main difference of the coded problem from the uncoded PIR (i.e., repetition-coded counterpart) is that in order to exploit the undesired coded bits in the form of side information, the interference needs to be decoded first. Note that we are not interested in decoding the individual components of each term of the sum, but rather the components of the \textit{aligned sum}. To perform this, we group each $K$ undesired equations to be from the same rows. In this case, we have $K$ linearly independent equations that can be uniquely solved, and hence the corresponding row of the interfering messages is decoded due to (\ref{decodeK}). Therefore, this generates $N\binom{M-1}{i}K^{M-(i+1)}(N-K)^{i-1}$ side information equations in the form of $i$ term sums. 
	
	\item\textit{Exploiting side information:} The side information generated in the previous step can be exploited in the $(i+1)$th round within the remaining $N-K$ databases that did not participate in generating them. The side information is used in $i+1$ term sum that includes the desired message as one of the terms. Since side information is successfully decoded, it can be canceled from these equations to leave desired coded bits. Hence, we can download $N\binom{M-1}{i}K^{M-(i+1)}(N-K)^{i}$ extra desired coded bits.
	\item Repeat steps 4, 5, 6 after setting $i=i+1$ until $i=M-1$.
	\item \textit{Decoding the desired message:} Till this point the scheme has downloaded one bit from each row of the desired message. To reliably decode the desired message, the scheme (precisely steps 2-7) is repeated $K$ times. We repeat the scheme exactly except for shifting the order of databases circularly at each repetition for the desired coded bits. Note that the chosen indices for the desired message is the same up to circular shift at each repetition, however we download new undesired coded bits at each repetition.  This creates $K$ different equations for each row of the message and hence decodable.
	\item \textit{Shuffling the order of queries:} Since all databases know the retrieval scheme, every database can identify the desired message by observing the first query only. By shuffling the order of queries uniformly, all possible queries can be made equally likely regardless of the message index. This guarantees the privacy.   	
\end{enumerate}
\subsection{Calculation of the Achievable Rate}
From the described scheme, we note that other than the initial download of $NK^{M-1}$ coded desired bits, at each round the scheme downloads $N\binom{M-1}{i}K^{M-(i+1)}(N-K)^{i}$ desired equations and $N\binom{M-1}{i}K^{M-i}(N-K)^{i-1}$ undesired equations. Hence, the total number of desired equations is $KN\sum_{i=0}^{M-1} \binom{M-1}{i} K^{M-1-i} (N-K)^i$, and the total number of undesired equations is $KN\sum_{i=1}^{M-1} \binom{M-1}{i} K^{M-i} (N-K)^{i-1}$ along the $K$ repetitions of the scheme. The achievable rate is,
\begin{align}
\frac{1}{R}&=1+\frac{\text{total undesired equations}}{\text{total desired equations}} \\
&=1+\frac{\sum_{i=1}^{M-1} \binom{M-1}{i} K^{M-i} (N-K)^{i-1}}{ \sum_{i=0}^{M-1} \binom{M-1}{i} K^{M-1-i} (N-K)^i} \\
&=1+\frac{\frac{K}{N-K}\sum_{i=1}^{M-1} \binom{M-1}{i} K^{M-1-i} (N-K)^{i}}{N^{M-1}}  \\
&=1+\frac{\frac{K}{N-K}\left(\sum_{i=0}^{M-1} \binom{M-1}{i} K^{M-1-i} (N-K)^i-K^{M-1}\right)}{N^{M-1}} \\
&=1+\frac{\frac{K}{N-K}\left(N^{M-1}-K^{M-1}\right)}{N^{M-1}} \\
&=1+\frac{K}{N-K}\left(1-R_c^{M-1}\right) \\
&=\frac{N-KR_c^{M-1}}{N-K} \\
&=\frac{1-R_c^M}{1-R_c}
\end{align}
Hence, $R=\frac{1-R_c}{1-R_c^M}$. Note that if $K=1$, our achievable scheme reduces to the one presented in \cite{JafarPIR}. We note that our scheme inherits all the properties of the scheme in \cite{JafarPIR}, in particular, its optimality over any subset of messages.
\section{Examples}\label{sec:examples}
In this section, we give two explicit examples for our scheme. Without loss of generality, we assume that the desired message is $W_1$.
\subsection{(5,3) Code with $M=2$}
Initially, sub-indices of all messages are randomly and independently interleaved. For this case, we will have $M=2$ rounds and then $K=3$ repetitions; see Table~\ref{code(5,3)}. We begin round one by downloading $K^{M-1}=3$ coded bits for the desired message (message $W_1$) from every database, e.g., we download $\bh_1^T \bx_1^{[1]},\bh_1^T \bx_2^{[1]},\bh_1^T \bx_3^{[1]}$ from database 1, and similarly for databases 2-5 by database symmetry. By message symmetry, we download another $3$ coded bits from $W_2$ from each database. Note that for the undesired message, we group every $K=3$ databases to download from the same row, e.g., we download $\bh_1^T \bx_1^{[2]},\bh_2^T \bx_1^{[2]},\bh_3^T \bx_1^{[2]}$ from databases 1-3, $\bh_4^T \bx_2^{[2]},\bh_5^T \bx_2^{[2]},\bh_1^T \bx_2^{[2]}$ from databases 4,5,1, and similarly for the remaining databases. By downloading 3 linearly independent equations for every row, we solve for the interference generated by $W_2$ and create $5$ useful side information rows for round two, which are rows $\bx_1^{[2]}$ to $\bx_5^{[2]}$ from $W_2$. 

In round two, we download sums of the coded bits from $W_1, W_2$. Since each of the rows $\bx_1^{[2]}$ to $\bx_5^{[2]}$ is decoded from 3 databases, we can exploit these side information to download further coded bits from $W_1$ in the remaining $N-K=2$ databases that do not participate in decoding this row. For example, we use $\bx_1^{[2]}$ in databases 4,5 by downloading the sums $\bh_4^T (\bx_{19}^{[1]}+\bx_1^{[2]})$, and $\bh_5^T (\bx_{20}^{[1]}+\bx_1^{[2]})$ and similarly for the rows $\bx_2^{[2]}$ to $\bx_5^{[2]}$. This creates extra $10$ decodable equations in round two in the form of a sum of the two messages. At this point symmetry exists across databases and within messages, and all the interference from the undesired message $W_2$ is decoded and exploited. However, until this point, we downloaded one equation from every row of $W_1$. To reliably decode $W_1$, we need to repeat the previous steps a total of $K=3$ times by shifting the starting database in circular pattern, e.g., in repetition 2, we download new equations for the rows $\bx_1^{[1]},\bx_2^{[1]},\bx_3^{[1]}$ from database 2 instead of database 1 in repetition 1, and $\bx_4^{[1]},\bx_5^{[1]},\bx_6^{[1]}$ from database 3 instead of database 2, etc. As a final step, we shuffle the order of the queries to preclude the databases from identifying the message index from the index of the first downloaded bit. 

Since we download symmetric amount of $W_1,W_2$ from each database and their indices are randomly chosen, privacy constraint is satisfied. Since vectors $\bx_i^{[2]}, i \in \{1, \cdots, 5\}$ are downloaded from $K$ databases, their interference is completely decoded. Hence, they can be canceled from round two. Finally, we repeat the scheme 3 times with circular shifts, every desired row is received from $K$ different databases and hence reliably decoded. The explicit query table  is shown in Table~\ref{code(5,3)}. The retrieval rate in this case is $R=\frac{75}{120}=\frac{5}{8}=\frac{1-\frac{3}{5}}{1-(\frac{3}{5})^2}$ .
\begin{table}[]
	\centering
	\caption{PIR for code (5,3) and $M=2$}
	\label{code(5,3)}
	\begin{tabular}{|l|l|c|c|c|c|c|}
		\hline
		\multicolumn{2}{|l|}{}                                                     & DB1                                       & DB2                                       & DB3                                       & DB4                                       & DB5                                       \\ \hline
		\multirow{8}{*}{\rotatebox[origin=c]{90}{\parbox[c]{3cm}{\centering repetition 1}}} & \multirow{6}{*}{\rotatebox[origin=c]{90}{\parbox[c]{2cm}{\centering round 1}}}      & $\bh_1^T \bx_1^{[1]}$                     & $\bh_2^T \bx_4^{[1]}$                     & $\bh_3^T \bx_7^{[1]}$                     & $\bh_4^T \bx_{10}^{[1]}$                  & $\bh_5^T \bx_{13}^{[1]}$                  \\
		&                                            & $\bh_1^T \bx_2^{[1]}$                     & $\bh_2^T \bx_5^{[1]}$                     & $\bh_3^T \bx_8^{[1]}$                     & $\bh_4^T \bx_{11}^{[1]}$                  & $\bh_{5}^T \bx_{14}^{[1]}$                \\
		&                                            & $\bh_1^T \bx_{3}^{[1]}$                   & $\bh_2^T \bx_{6}^{[1]}$                   & $\bh_3^T \bx_{9}^{[1]}$                   & $\bh_4^T \bx_{12}^{[1]}$                  & $\bh_5^T \bx_{15}^{[1]}$                  \\
		&                                            & $\bh_1^T \bx_1^{[2]}$                     & $\bh_2^T \bx_1^{[2]}$                     & $\bh_3^T \bx_1^{[2]}$                     & $\bh_4^T \bx_2^{[2]}$                     & $\bh_5^T \bx_2^{[2]}$                     \\
		&                                            & $\bh_1^T \bx_2^{[2]}$                     & $\bh_2^T \bx_3^{[2]}$                     & $\bh_3^T \bx_3^{[2]}$                     & $\bh_4^T \bx_3^{[2]}$                     & $\bh_{5}^T \bx_{4}^{[2]}$                 \\
		&                                            & $\bh_1^T \bx_{4}^{[2]}$                   & $\bh_2^T \bx_{4}^{[2]}$                   & $\bh_3^T \bx_{5}^{[2]}$                   & $\bh_4^T \bx_{5}^{[2]}$                   & $\bh_5^T \bx_{5}^{[2]}$                   \\ \cline{2-7} 
		\rule{0pt}{3ex}
		& \multirow{2}{*}{\rotatebox[origin=c]{90}{\parbox[c]{1cm}{\centering round 2}}} & $\bh_1^T (\bx_{16}^{[1]}+\bx_3^{[2]})$    & $\bh_2^T (\bx_{17}^{[1]}+\bx_2^{[2]})$    & $\bh_3^T (\bx_{18}^{[1]}+\bx_2^{[2]})$    & $\bh_4^T (\bx_{19}^{[1]}+\bx_1^{[2]})$    & $\bh_5^T (\bx_{20}^{[1]}+\bx_1^{[2]})$    \\
		&                                            & $\bh_1^T (\bx_{21}^{[1]}+\bx_5^{[2]})$    & $\bh_2^T (\bx_{22}^{[1]}+\bx_5^{[2]})$    & $\bh_3^T (\bx_{23}^{[1]}+\bx_4^{[2]})$    & $\bh_4^T (\bx_{24}^{[1]}+\bx_4^{[2]})$    & $\bh_5^T (\bx_{25}^{[1]}+\bx_3^{[2]})$    \\ \hline
		\multirow{8}{*}{\rotatebox[origin=c]{90}{\parbox[c]{3cm}{\centering repetition 2}}} & \multirow{6}{*}{\rotatebox[origin=c]{90}{\parbox[c]{2cm}{\centering round 1}}}        & $\bh_1^T \bx_{13}^{[1]}$                  & $\bh_2^T \bx_1^{[1]}$                     & $\bh_3^T \bx_4^{[1]}$                     & $\bh_4^T \bx_{7}^{[1]}$                   & $\bh_5^T \bx_{10}^{[1]}$                  \\
		&                                            & $\bh_1^T \bx_{14}^{[1]}$                  & $\bh_2^T \bx_2^{[1]}$                     & $\bh_3^T \bx_5^{[1]}$                     & $\bh_4^T \bx_{8}^{[1]}$                   & $\bh_{5}^T \bx_{11}^{[1]}$                \\
		&                                            & $\bh_1^T \bx_{15}^{[1]}$                  & $\bh_2^T \bx_{3}^{[1]}$                   & $\bh_3^T \bx_{6}^{[1]}$                   & $\bh_4^T \bx_{9}^{[1]}$                   & $\bh_5^T \bx_{12}^{[1]}$                  \\
		&                                            & $\bh_1^T \bx_6^{[2]}$                     & $\bh_2^T \bx_6^{[2]}$                     & $\bh_3^T \bx_6^{[2]}$                     & $\bh_4^T \bx_7^{[2]}$                     & $\bh_5^T \bx_7^{[2]}$                     \\
		&                                            & $\bh_1^T \bx_7^{[2]}$                     & $\bh_2^T \bx_8^{[2]}$                     & $\bh_3^T \bx_8^{[2]}$                     & $\bh_4^T \bx_8^{[2]}$                     & $\bh_{5}^T \bx_{9}^{[2]}$                 \\
		&                                            & $\bh_1^T \bx_{9}^{[2]}$                   & $\bh_2^T \bx_{9}^{[2]}$                   & $\bh_3^T \bx_{10}^{[2]}$                  & $\bh_4^T \bx_{10}^{[2]}$                  & $\bh_5^T \bx_{10}^{[2]}$                  \\ \cline{2-7} 
		\rule{0pt}{3ex}
		& \multirow{2}{*}{\rotatebox[origin=c]{90}{\parbox[c]{1cm}{\centering round 2}}} & $\bh_1^T (\bx_{20}^{[1]}+\bx_8^{[2]})$    & $\bh_2^T (\bx_{16}^{[1]}+\bx_7^{[2]})$    & $\bh_3^T (\bx_{17}^{[1]}+\bx_7^{[2]})$    & $\bh_4^T (\bx_{18}^{[1]}+\bx_6^{[2]})$    & $\bh_5^T (\bx_{19}^{[1]}+\bx_6^{[2]})$    \\
		&                                            & $\bh_1^T (\bx_{25}^{[1]}+\bx_{10}^{[2]})$ & $\bh_2^T (\bx_{21}^{[1]}+\bx_{10}^{[2]})$ & $\bh_3^T (\bx_{22}^{[1]}+\bx_9^{[2]})$    & $\bh_4^T (\bx_{23}^{[1]}+\bx_9^{[2]})$    & $\bh_5^T (\bx_{24}^{[1]}+\bx_8^{[2]})$    \\ \hline
		\multirow{8}{*}{\rotatebox[origin=c]{90}{\parbox[c]{3cm}{\centering repetition 3}}}& \multirow{6}{*}{\rotatebox[origin=c]{90}{\parbox[c]{2cm}{\centering round 1}}}       & $\bh_1^T \bx_{10}^{[1]}$                  & $\bh_2^T \bx_{13}^{[1]}$                  & $\bh_3^T \bx_1^{[1]}$                     & $\bh_4^T \bx_{4}^{[1]}$                   & $\bh_5^T \bx_{7}^{[1]}$                   \\
		&                                            & $\bh_1^T \bx_{11}^{[1]}$                  & $\bh_2^T \bx_{14}^{[1]}$                  & $\bh_3^T \bx_2^{[1]}$                     & $\bh_4^T \bx_{5}^{[1]}$                   & $\bh_{5}^T \bx_{8}^{[1]}$                 \\
		&                                            & $\bh_1^T \bx_{12}^{[1]}$                  & $\bh_2^T \bx_{15}^{[1]}$                  & $\bh_3^T \bx_{3}^{[1]}$                   & $\bh_4^T \bx_{6}^{[1]}$                   & $\bh_5^T \bx_{9}^{[1]}$                   \\
		&                                            & $\bh_1^T \bx_{11}^{[2]}$                  & $\bh_2^T \bx_{11}^{[2]}$                  & $\bh_3^T \bx_{11}^{[2]}$                  & $\bh_4^T \bx_{12}^{[2]}$                  & $\bh_5^T \bx_{12}^{[2]}$                  \\
		&                                            & $\bh_1^T \bx_{12}^{[2]}$                  & $\bh_2^T \bx_{13}^{[2]}$                  & $\bh_3^T \bx_{13}^{[2]}$                  & $\bh_4^T \bx_{13}^{[2]}$                  & $\bh_{5}^T \bx_{14}^{[2]}$                \\
		&                                            & $\bh_1^T \bx_{14}^{[2]}$                  & $\bh_2^T \bx_{14}^{[2]}$                  & $\bh_{13}^T \bx_{15}^{[2]}$               & $\bh_4^T \bx_{15}^{[2]}$                  & $\bh_5^T \bx_{15}^{[2]}$                  \\ \cline{2-7} 
		\rule{0pt}{3ex}
		& \multirow{2}{*}{\rotatebox[origin=c]{90}{\parbox[c]{1cm}{\centering round 2}}} & $\bh_1^T (\bx_{19}^{[1]}+\bx_{13}^{[2]})$ & $\bh_2^T (\bx_{20}^{[1]}+\bx_{12}^{[2]})$ & $\bh_3^T (\bx_{16}^{[1]}+\bx_{12}^{[2]})$ & $\bh_4^T (\bx_{17}^{[1]}+\bx_{11}^{[2]})$ & $\bh_5^T (\bx_{18}^{[1]}+\bx_{11}^{[2]})$ \\
		&                                            & $\bh_1^T (\bx_{24}^{[1]}+\bx_{15}^{[2]})$ & $\bh_2^T (\bx_{25}^{[1]}+\bx_{15}^{[2]})$ & $\bh_3^T (\bx_{21}^{[1]}+\bx_{14}^{[2]})$ & $\bh_4^T (\bx_{22}^{[1]}+\bx_{14}^{[2]})$ & $\bh_5^T (\bx_{23}^{[1]}+\bx_{13}^{[2]})$ \\ \hline
	\end{tabular}
\end{table}
\subsection{(3,2) Code with $M=3$}
As in the previous example, the messages are randomly and  independently interleaved. For this case, the scheme is completed in $M=3$ rounds and then repeated for $K=2$ repetitions, see Table~\ref{code(3,2)}. In the first round, we download $K^{M-1}=4$ coded bits for $W_1$ from each database, e.g., $\bh_1^T\bx_i^{[1]}, i\in \{1, \cdots, 4\}$ from the first database. Similarly, we download one equation from the rows $\bx_1^{[1]}$ to $\bx_{12}^{[1]}$ by applying the database symmetry. We apply message symmetry to download $N \binom{M-1}{1} K^{M-1}=24$ undesired coded bits from $W_2,W_3$. Every $2$ coded bits from the undesired bits are grouped together to generate single solved side information vector, e.g., we download as $\bh_1^T \bx_1^{[2]}, \bh_2^T \bx_2^{[2]}$ from databases 1,2, $\bh_3^T \bx_2^{[2]}, \bh_1^T \bx_1^{[2]}$ from databases 3,1, and similarly for rows $\bx_1^{[m]}$ to $\bx_6^{[m]}$ where $m =2,3$. Hence, we have $N \binom{M-1}{1} K^{M-2}=12$ side information to be used in round two.

In round two, we download sums of every two messages. We exploit the generated side information within the $N-K=1$ remaining database that does not participate in generating them. For example, we decoded $\bx_1^{[2]}$ by downloading equations from databases 1,2, then we use $\bx_1^{[2]}$ in database 3 by downloading the sum $\bh_3(\bx_{15}^{[1]}+\bx_1^{[2]})$.  Hence, we can  download $N\binom{M-1}{1} K^{M-2}(N-K)=12$ new coded bits of $W_1$ by using every decoded side information in a sum of $W_1$ with one of $W_2$ or $W_3$. These bits are reliably decoded, since the generated side information can be canceled from the downloaded equation. It remains to add sums of $W_2$ and $W_3$ to ensure the privacy. Therefore, we download $N \binom{M-1}{2} K^{M-2}(N-K)=6$ undesired equations, that will be grouped further to form $N \binom{M-1}{2} K^{M-3}(N-K)=3$ solved side information equations in the form of sums of $W_2$ and $W_3$. As an example, we download $\bh_1^T(\bx_7^{[2]}+\bx_7^{[3]}), \bh_2^T(\bx_7^{[2]}+\bx_7^{[3]})$ from databases 1,2. In this case the interference from the rows $\bx_7^{[2]}+\bx_7^{[3]}$ is decoded. Note that we do not solve for the individual $\bx_7^{[2]}$ or $\bx_7^{[3]}$ but we \textit{align} them in the same subspace, and solve for their sum. 

In round three, we use the newly generated side information, e.g., $\bx_7^{[2]}+\bx_7^{[3]}$,  to download extra $N \binom{M-1}{2} K^{M-3}(N-K)^2=3$ desired coded bits in the form of sum of three terms, e.g., $\bh_3^T(\bx_{27}^{[1]}+\bx_7^{[2]}+\bx_7^{[3]})$. Finally, the previous steps are repeated $K=2$ times to reliably decode $W_1$ and the queries are shuffled for privacy. The retrieval rate in this case is $R=\frac{54}{114}=\frac{9}{19}=\frac{1-\frac{2}{3}}{1-(\frac{2}{3})^3}$. The explicit query structure is shown in Table~\ref{code(3,2)}.
\begin{table}[]
	\centering
	\caption{PIR for code (3,2) and $M=3$}
	\label{code(3,2)}
	\begin{tabular}{|l|l|c|c|c|}
		\hline
		\multicolumn{2}{|l|}{}                  & DB1                                                     & DB2                                                     & DB3                                                     \\ \hline
		\multirow{19}{*}{{\rotatebox[origin=c]{90}{\parbox[c]{3cm}{\centering repetition 1}}}} & \multirow{12}{*}{\rotatebox[origin=c]{90}{\parbox[c]{2cm}{\centering round 1}}} & $\bh_1^T \bx_1^{[1]}$                                   & $\bh_2^T \bx_5^{[1]}$                                   & $\bh_3^T \bx_9^{[1]}$                                   \\
		&                    & $\bh_1^T \bx_2^{[1]}$                                   & $\bh_2^T \bx_6^{[1]}$                                   & $\bh_3^T \bx_{10}^{[1]}$                                \\
		&                    & $\bh_1^T \bx_3^{[1]}$                                   & $\bh_2^T \bx_7^{[1]}$                                   & $\bh_3^T \bx_{11}^{[1]}$                                \\
		&                    & $\bh_1^T \bx_{4}^{[1]}$                                 & $\bh_2^T \bx_{8}^{[1]}$                                 & $\bh_3^T \bx_{12}^{[1]}$                                \\
		&                    & $\bh_1^T \bx_1^{[2]}$                                   & $\bh_2^T \bx_1^{[2]}$                                   & $\bh_3^T \bx_2^{[2]}$                                   \\
		&                    & $\bh_1^T \bx_2^{[2]}$                                   & $\bh_2^T \bx_3^{[2]}$                                   & $\bh_3^T \bx_3^{[2]}$                                   \\
		&                    & $\bh_1^T \bx_4^{[2]}$                                   & $\bh_2^T \bx_4^{[2]}$                                   & $\bh_3^T \bx_5^{[2]}$                                   \\
		&                    & $\bh_1^T \bx_{5}^{[2]}$                                 & $\bh_2^T \bx_{6}^{[2]}$                                 & $\bh_3^T \bx_{6}^{[2]}$                                 \\
		&                    & $\bh_1^T \bx_1^{[3]}$                                   & $\bh_2^T \bx_1^{[3]}$                                   & $\bh_3^T \bx_2^{[3]}$                                   \\
		&                    & $\bh_1^T \bx_2^{[3]}$                                   & $\bh_2^T \bx_3^{[3]}$                                   & $\bh_3^T \bx_3^{[3]}$                                   \\
		&                    & $\bh_1^T \bx_4^{[3]}$                                   & $\bh_2^T \bx_4^{[3]}$                                   & $\bh_3^T \bx_5^{[3]}$                                   \\
		&                    & $\bh_1^T \bx_{5}^{[3]}$                                 & $\bh_2^T \bx_{6}^{[3]}$                                 & $\bh_3^T \bx_{6}^{[3]}$                                 \\ \cline{2-5} 
		& \multirow{6}{*}{\rotatebox[origin=c]{90}{\parbox[c]{2cm}{\centering round 2}}}  & $\bh_1^T (\bx_{13}^{[1]}+\bx_3^{[2]})$                  & $\bh_2^T (\bx_{14}^{[1]}+\bx_2^{[2]})$                  & $\bh_3^T (\bx_{15}^{[1]}+\bx_1^{[2]})$                  \\
		&                    & $\bh_1^T (\bx_{16}^{[1]}+\bx_3^{[3]})$                  & $\bh_2^T (\bx_{17}^{[1]}+\bx_2^{[3]})$                  & $\bh_3^T (\bx_{18}^{[1]}+\bx_1^{[3]})$                  \\
		&                    & $\bh_1^T (\bx_{7}^{[2]}+\bx_7^{[3]})$                   & $\bh_2^T (\bx_{7}^{[2]}+\bx_7^{[3]})$                   & $\bh_3^T (\bx_{8}^{[2]}+\bx_8^{[3]})$                   \\
		&                    & $\bh_1^T (\bx_{19}^{[1]}+\bx_6^{[2]})$                  & $\bh_2^T (\bx_{20}^{[1]}+\bx_5^{[2]})$                  & $\bh_3^T (\bx_{21}^{[1]}+\bx_4^{[2]})$                  \\
		&                    & $\bh_1^T (\bx_{22}^{[1]}+\bx_6^{[3]})$                  & $\bh_2^T (\bx_{23}^{[1]}+\bx_5^{[3]})$                  & $\bh_3^T (\bx_{24}^{[1]}+\bx_4^{[3]})$                  \\
		&                    & $\bh_1^T (\bx_{8}^{[2]}+\bx_8^{[3]})$                   & $\bh_2^T (\bx_{9}^{[2]}+\bx_9^{[3]})$                   & $\bh_3^T (\bx_{9}^{[2]}+\bx_9^{[3]})$                   \\ \cline{2-5}
		\rule{0pt}{3ex} 
		& \multirow{1}{*}{\rotatebox[origin=c]{90}{\parbox[c]{0.5cm}{\centering rd. 3}}}                   & $\bh_1^T(\bx_{25}^{[1]}+\bx_{9}^{[2]}+\bx_9^{[3]})$     & $\bh_2^T(\bx_{26}^{[1]}+\bx_{8}^{[2]}+\bx_8^{[3]})$     & $\bh_3^T(\bx_{27}^{[1]}+\bx_{7}^{[2]}+\bx_7^{[3]})$     \\ \hline
		\multirow{19}{*}{\rotatebox[origin=c]{90}{\parbox[c]{3cm}{\centering repetition 2}}} & \multirow{12}{*}{\rotatebox[origin=c]{90}{\parbox[c]{2cm}{\centering round 1}}} & $\bh_1^T \bx_9^{[1]}$                                   & $\bh_2^T \bx_1^{[1]}$                                   & $\bh_3^T \bx_5^{[1]}$                                   \\
		&                    & $\bh_1^T \bx_{10}^{[1]}$                                & $\bh_2^T \bx_2^{[1]}$                                   & $\bh_3^T \bx_{6}^{[1]}$                                 \\
		&                    & $\bh_1^T \bx_{11}^{[1]}$                                & $\bh_2^T \bx_3^{[1]}$                                   & $\bh_3^T \bx_{7}^{[1]}$                                 \\
		&                    & $\bh_1^T \bx_{12}^{[1]}$                                & $\bh_2^T \bx_{4}^{[1]}$                                 & $\bh_3^T \bx_{8}^{[1]}$                                 \\
		&                    & $\bh_1^T \bx_{10}^{[2]}$                                & $\bh_2^T \bx_{10}^{[2]}$                                & $\bh_3^T \bx_{11}^{[2]}$                                \\
		&                    & $\bh_1^T \bx_{11}^{[2]}$                                & $\bh_2^T \bx_{12}^{[2]}$                                & $\bh_3^T \bx_{12}^{[2]}$                                \\
		&                    & $\bh_1^T \bx_{13}^{[2]}$                                & $\bh_2^T \bx_{13}^{[2]}$                                & $\bh_3^T \bx_{14}^{[2]}$                                \\
		&                    & $\bh_1^T \bx_{14}^{[2]}$                                & $\bh_2^T \bx_{15}^{[2]}$                                & $\bh_3^T \bx_{15}^{[2]}$                                \\
		&                    & $\bh_1^T \bx_{10}^{[3]}$                                & $\bh_2^T \bx_{10}^{[3]}$                                & $\bh_3^T \bx_{11}^{[3]}$                                \\
		&                    & $\bh_1^T \bx_{11}^{[3]}$                                & $\bh_2^T \bx_{12}^{[3]}$                                & $\bh_3^T \bx_{12}^{[3]}$                                \\
		&                    & $\bh_1^T \bx_{13}^{[3]}$                                & $\bh_2^T \bx_{13}^{[3]}$                                & $\bh_3^T \bx_{14}^{[3]}$                                \\
		&                    & $\bh_1^T \bx_{14}^{[3]}$                                & $\bh_2^T \bx_{15}^{[3]}$                                & $\bh_3^T \bx_{15}^{[3]}$                                \\ \cline{2-5} 
		& \multirow{6}{*}{\rotatebox[origin=c]{90}{\parbox[c]{2cm}{\centering round 2}}}  & $\bh_1^T (\bx_{15}^{[1]}+\bx_{12}^{[2]})$               & $\bh_2^T (\bx_{13}^{[1]}+\bx_{11}^{[2]})$               & $\bh_3^T (\bx_{14}^{[1]}+\bx_{10}^{[2]})$               \\
		&                    & $\bh_1^T (\bx_{18}^{[1]}+\bx_{12}^{[3]})$               & $\bh_2^T (\bx_{16}^{[1]}+\bx_{11}^{[3]})$               & $\bh_3^T (\bx_{17}^{[1]}+\bx_{10}^{[3]})$               \\
		&                    & $\bh_1^T (\bx_{16}^{[2]}+\bx_{16}^{[3]})$               & $\bh_2^T (\bx_{16}^{[2]}+\bx_{16}^{[3]})$               & $\bh_3^T (\bx_{17}^{[2]}+\bx_{17}^{[3]})$               \\
		&                    & $\bh_1^T (\bx_{21}^{[1]}+\bx_{15}^{[2]})$               & $\bh_2^T (\bx_{19}^{[1]}+\bx_{14}^{[2]})$               & $\bh_3^T (\bx_{20}^{[1]}+\bx_{13}^{[2]})$               \\
		&                    & $\bh_1^T (\bx_{24}^{[1]}+\bx_{15}^{[3]})$               & $\bh_2^T (\bx_{22}^{[1]}+\bx_{14}^{[3]})$               & $\bh_3^T (\bx_{23}^{[1]}+\bx_{13}^{[3]})$               \\
		&                    & $\bh_1^T (\bx_{17}^{[2]}+\bx_{17}^{[3]})$               & $\bh_2^T (\bx_{18}^{[2]}+\bx_{18}^{[3]})$               & $\bh_3^T (\bx_{18}^{[2]}+\bx_{18}^{[3]})$               \\ \cline{2-5} 
			\rule{0pt}{3ex} 
			& \multirow{1}{*}{\rotatebox[origin=c]{90}{\parbox[c]{0.5cm}{\centering rd. 3}}}                 & $\bh_1^T(\bx_{27}^{[1]}+\bx_{18}^{[2]}+\bx_{18}^{[3]})$ & $\bh_2^T(\bx_{25}^{[1]}+\bx_{17}^{[2]}+\bx_{17}^{[3]})$ & $\bh_3^T(\bx_{26}^{[1]}+\bx_{16}^{[2]}+\bx_{16}^{[3]})$ \\ \hline
	\end{tabular}
\end{table}
\section{Converse Proof}
\subsection{Notations and Simplifications}
We follow the notations and simplifications of \cite{JafarPIR}. We define, 
\begin{align}
\cq &\triangleq \{Q_n^{[m]}:m \in \{1, \cdots, M\},\quad n \in \{1,\cdots, N\}\} \\
A_{n_1:n_2}^{[m]}&\triangleq\{A_{n_1}^{[m]}, \cdots,A_{n_2}^{[m]}\}, \quad n_1 \leq n_2,\quad n_1,n_2 \in \{1, \cdots, N\}
\end{align}
We use $(n_1:n_2) \mod N$ to denote the circular indices from $n_1$ to $n_2$, i.e., if $n \geq N$, then $n$ is replaced by $(n\,\text{mod}\,N)$. Without loss of generality, we can make the following simplifications \cite{JafarPIR}:
\begin{enumerate}
\item We can assume that the PIR scheme is symmetric. This can be assumed without loss of generality, since for any asymmetric PIR scheme, one can construct an equivalent symmetric retrieval scheme that has the same retrieval rate by replicating all permutations of databases and messages with appropriate time sharing.
\item We can invoke the non-colluding privacy constraint by fixing the query to one database to be the same irrespective of the desired message, i.e., $Q_n^{[m]}=Q_n, m \in \{1, \cdots, M\}$ for some $n \in \{1, \cdots, N\}$. This implies that $A_n^{[m]}=A_n, m \in \{1, \cdots, M\}$. This simplification is without loss of generality, since the queries are independent of the desired message index. Note that the index of this database can be chosen arbitrarily, and hence without loss of generality, we choose it to be the first database, i.e., $A_1^{[m]}=A_1, \forall m$.  
\end{enumerate}  
We first state the following lemma whose proof can be found in \cite[Lemma~1]{JafarPIR}. 
\begin{lemma}[Symmetry \cite{JafarPIR}]\label{lemma1}
	Without loss of generality, we have
	\begin{align}
	H(A_n^{[1]}|W_2, \cdots, W_M,\cq)&=H(A_1^{[1]}|W_2, \cdots, W_M,\cq), \quad n \in \{1, \cdots, N\} \\
	H(A_1^{[1]}|W_2, \cdots, W_M,\cq)&\geq \frac{H(W_1)}{N}=\frac{L}{N} \\
	\label{symmetry3} H(A_1|\cq)&=H(A_n^{[m]}|\cq), \quad \forall m \in \{1, \cdots, M\}, n \in \{1, \cdots, N\} 
	\end{align}
\end{lemma}

We note that the equality in (\ref{symmetry3}) remains true if the answer strings are conditioned on any subset $W_\mathcal{S}=\{W_i: i \in \mathcal{S}\}$ of messages, i.e.,
\begin{align}\label{symmetryCond}
H(A_1|W_\mathcal{S},\cq)=H(A_n^{[m]}|W_\mathcal{S},\cq), \quad \forall m,n
\end{align}
because otherwise the $n$th database can break the privacy requirement by conditioning the answer strings on $W_\mathcal{S}$ before responding to the user, and from the difference in lengths, the database can infer some information about the desired message index. 
\begin{lemma}[Independence of answers of any $K$ databases]\label{lemma2}
	For any set $\mathcal{K}$ of databases such that $|\mathcal{K}|=K$,
	\begin{align} \label{indepH}
	H(A_{\mathcal{K}}^{[1]}|\cq)=KH(A_1^{[1]}|\cq)
	\end{align}
Furthermore, (\ref{indepH}) is true if conditioned on any subset of messages $W_\mathcal{S}$, i.e.,
\begin{align}\label{indepHS}
H(A_{\mathcal{K}}^{[1]}|W_\mathcal{S},\cq)=KH(A_1^{[1]}|W_\mathcal{S},\cq)
\end{align} 
\end{lemma}
\begin{Proof}
	Consider a set of databases $\mathcal{K}$ such that $|\mathcal{K}|=K$. We prove first the statistical independence between the vectors $\{\by_n, n \in \mathcal{K}\}$ where $\by_n$ represents the contents of the $n$th database. The contents of set $\mathcal{K}$ of databases can be written as
	\begin{align}
	[\by_n, \: n \in \mathcal{K}]=\begin{bmatrix} W_1 \\\vdots\\ W_M
	\end{bmatrix} [\bh_n, \: n \in \mathcal{K}]=\begin{bmatrix} W_1 \\\vdots\\ W_M
	\end{bmatrix} \mathbf{H}_{\mathcal{K}}
	\end{align}
	where $\mathbf{H}_{\mathcal{K}}= [\bh_n, \: n \in \mathcal{K}]$ is a $\mathbb{F}_q^{K \times K}$ matrix. By construction of the distributed storage code, the matrix $\mathbf{H}_{\mathcal{K}}$ is an invertible matrix. Using \cite[Lemma~1]{JafarColluding} and the fact that elements of the messages are chosen independently and uniformly over $\mathbb{F}_q^{\tilde{L} \times K}$, we conclude that
	\begin{align}
	[\by_n, \: n \in \mathcal{K}]=\begin{bmatrix} W_1 \\\vdots\\ W_M
	\end{bmatrix} \mathbf{H}_{\mathcal{K}} \sim \begin{bmatrix} W_1 \\\vdots\\ W_M
	\end{bmatrix}
	\end{align} 
	where $A \sim B$ denotes that random variables $A$ and $B$ are identically distributed. Therefore, the contents of the databases are statistically equivalent to the messages. Hence, the columns of $[\by_n, \: n \in \mathcal{K}]$ are statistically independent since the elements of the messages are independent.
	
	Since $A_n^{[1]},  n \in \mathcal{K}$ are deterministic functions of $(\by_n,\cq)$, $\{A_n^{[1]}: n \in \mathcal{K}\}$ are statistically independent as they are deterministic functions of independent random variables. Due to the symmetry in Lemma~\ref{lemma1}, we have  $H(A_{\mathcal{K}}^{[1]}|\cq)=KH(A_1^{[1]}|\cq)$. We note that since coding is applied on individual messages, conditioning on any subset of messages $W_\mathcal{S}$ with $|W_\mathcal{S}|=S$ is equivalent to reducing the problem to storing $M-S$ independent messages instead of $M$ messages. Hence, the statistical independence argument in (\ref{indepHS}) follows as before.
\end{Proof}
\subsection{Converse Proof of the Case $M=2$}
\begin{lemma}[Interference lower bound]\label{lemma3}
	For the case $M=2$, the uncertainty on the interference from $W_2$ in the answers $A_{1:N}^{[1]}$ is lower bounded as,
	\begin{align}
	H(A_{1:N}^{[1]}|W_1,\cq) \geq \frac{K}{N} H(W_1)=\frac{KL}{N}
	\end{align}
\end{lemma}
\begin{Proof}
	For some set $\mathcal{K} \subset \{1,\cdots, N\}$ of databases such that $|\mathcal{K}|=K$, we can write,
	\begin{align}
	H(A_{1:N}^{[1]}|W_1,\cq) 
	&\geq H(A_{\mathcal{K}}^{[1]}|W_1,\cq) \\
	&\label{Lemma2result}=KH(A_1^{[1]}|W_1,\cq) \\
	&\label{answerA1}=KH(A_1^{[1]}|W_2,\cq) \\
	&\label{lemma1result}\geq K\frac{H(W_1)}{N}= \frac{KL}{N}
	\end{align}
	where (\ref{Lemma2result}) follows from Lemma~\ref{lemma2}, (\ref{answerA1}) follows from the privacy constraint since if $H(A_1^{[1]}|W_1,\cq) \neq H(A_1^{[1]}|W_2,\cq)$, database 1 can differentiate between the messages based on conditioning the answer strings on $W_1, W_2$, respectively, and (\ref{lemma1result}) follows from Lemma~\ref{lemma1}.
\end{Proof}

Now, we are ready to derive the converse proof for the case $M=2$,
\begin{align}
L&=H(W_1) \\
&\label{indep}=H(W_1|\cq) \\
&\label{decode}=H(W_1|\cq)-H(W_1|A_{1:N}^{[1]},\cq)\\
&=I(W_1;A_{1:N}^{[1]}|\cq)\\
&=H(A_{1:N}^{[1]}|\cq)-H(A_{1:N}^{[1]}|W_1,\cq)\\
&\label{Lemma3result}\leq H(A_{1:N}^{[1]}|\cq)-\frac{KL}{N}\\
&\label{conditioning}\leq \sum_{n=1}^N  H(A_n^{[1]}|\cq)-\frac{KL}{N}
\end{align}
where (\ref{indep}) follows from the independence of the queries and the messages, (\ref{decode}) follows from the reliability constraint (\ref{reliability}) for $W_1$, (\ref{Lemma3result}) follows from Lemma~\ref{lemma3}, and (\ref{conditioning}) follows from the fact that conditioning does not increase entropy. Hence, using Lemma~\ref{lemma1}, 
\begin{align}\label{inductM=2}
NH(A_1|\cq) \geq L\left(1+\frac{K}{N}\right)
\end{align}
Then, using (\ref{PIRrate}), the retrieval rate is upper bounded by, 
\begin{align}\label{upperM=2}
R=\frac{L}{\sum_{n=1}^N H(A_n^{[1]})} \leq \frac{L}{N H(A_1|\cq)} \leq \frac{1}{1+\frac{K}{N}}
\end{align}
\subsection{Converse Proof for $M \geq 3$}
We use a technique similar to that in \cite{JafarPIR}. In the sequel, we derive an inductive relation that can be used in addition to the base induction step of $M=2$ to obtain a matching upper bound for the achievable rate in (\ref{PIR_rate}). We need the following lemma which upper bounds the uncertainty on the answer strings after knowing one of the interference messages.
\begin{lemma}[Interference conditioning lemma] \label{conditioningLemma}
	The remaining uncertainty on the answer strings after conditioning on one of the interfering messages is upper bounded by,
	\begin{align}
	H(A_{1:N}^{[2]}|W_1,\cq) \leq \frac{N}{K}\left(NH(A_1|\cq)-L\right) 
	\end{align}
\end{lemma}
\begin{Proof}
We have
	\begin{align}
	H(&A_{1:N}^{[2]}|W_1,\cq) \notag\\
	&\leq \sum_{n=1}^N H(A_n^{[2]}|W_1,\cq) \\
	\label{ind5}			   &= \sum_{n=1}^N \frac{1}{K} \left(H(A_n^{[2]}|W_1,\cq)+\sum_{j=n+1\,\text{mod}\,N}^{n+K-1\,\text{mod}\, N}H(A_j^{[1]}|W_1,\cq)\right) \\
	\label{ind6}               &= \frac{1}{K} \sum_{n=1}^N  H(A_{(n+1:n+K-1)\,\text{mod}\,N}^{[1]},A_n^{[2]}|W_1,\cq) \\
	&\label{indAlign}\leq \frac{1}{K} \sum_{n=1}^N  H(A_{1:n-1}^{[1]},A_n^{[2]},A_{n+1:N}^{[1]}|W_1,\cq) \\
	&=\frac{1}{K} \sum_{n=1}^N \left(H(A_{1:n-1}^{[1]},A_n^{[2]},A_{n+1:N}^{[1]},W_1|\cq)-H(W_1|\cq)\right) \\
	\label{ind7}               &=\frac{1}{K} \sum_{n=1}^N \left(H(A_{1:n-1}^{[1]},A_n^{[2]},A_{n+1:N}^{[1]}|\cq)+H(W_1|A_{1:n-1}^{[1]},A_n^{[2]},A_{n+1:N}^{[1]},\cq)-H(W_1)\right) \\
	\label{ind8}               &\leq \frac{1}{K} \sum_{n=1}^N \left(NH(A_1|\cq)-H(W_1)\right) \\
	&=\frac{N}{K}\left(NH(A_1|\cq)-L\right)           
	\end{align}
where (\ref{ind5}) follows from the symmetry of the answer strings along databases and messages stated in (\ref{symmetryCond}), (\ref{ind6}) follows from the independence of any $K$ answer strings as shown in Lemma~\ref{lemma2}, (\ref{ind7}) follows from the independence between $W_1$ and the queries, and (\ref{ind8}) follows from the independence bound, symmetry in Lemma~\ref{lemma1}, and the decodability of $W_1$ given the answer strings $(A_{1:n-1}^{[1]},A_n^{[2]},A_{n+1:N}^{[1]})$.
\end{Proof}

From the proof, we can see that this lemma is crucial, since it captures the main aspects of the problem, namely: coding which appears in the independence of (\ref{ind6}), privacy in the form of the ability of fixing one of the answers for two different messages in (\ref{ind8}), and interference alignment in (\ref{indAlign}). 

Now, we are ready for the proof of the inductive relation for general $M$,
\begin{align}
\label{indstart}		ML=&H(W_1, \cdots, W_M|\cq) \\
\label{ind1}   =&H(W_1, \cdots, W_M|\cq)-H(W_1, \cdots, W_M|A_{1:N}^{[1]}, \cdots, A_{1:N}^{[M]},\cq)\\
			   =&I(A_{1:N}^{[1]}, \cdots, A_{1:N}^{[M]};W_1, \cdots, W_M|\cq) \\
\label{ind2}   =&H(A_{1:N}^{[1]}, \cdots, A_{1:N}^{[M]}|\cq) \\
			   =&H(A_{1:N}^{[1]}|\cq)+H(A_{1:N}^{[2]}, \cdots, A_{1:N}^{[M]}|A_{1:N}^{[1]},\cq)	\\
\label{ind3}			   =&H(A_{1:N}^{[1]}|\cq)+H(A_{1:N}^{[2]}, \cdots, A_{1:N}^{[M]}|A_{1:N}^{[1]},W_1,\cq)	\\
   =&H(A_{1:N}^{[1]}|\cq)+H(A_{1:N}^{[1]}, \cdots, A_{1:N}^{[M]}|W_1,\cq)-H(A_{1:N}^{[1]}|W_1,\cq)	\\
			   =&I(A_{1:N}^{[1]};W_1|\cq)+H(A_{1:N}^{[2]}|W_1,\cq)+H(A_{1:N}^{[1]},A_{1:N}^{[3]}, \cdots, A_{1:N}^{[M]}|A_{1:N}^{[2]},W_1,\cq)\\
\label{ind4}			   =&I(A_{1:N}^{[1]};W_1|\cq)+H(A_{1:N}^{[2]}|W_1,\cq)+H(A_{1:N}^{[1]},A_{1:N}^{[3]}, \cdots, A_{1:N}^{[M]}|A_{1:N}^{[2]},W_1,W_2,\cq)\\
			   =&I(A_{1:N}^{[1]};W_1|\cq)+H(A_{1:N}^{[2]}|W_1,\cq)+H(A_{1:N}^{[1]}, \cdots, A_{1:N}^{[M]}|W_1,W_2,\cq)\notag\\
			   &-H(A_{1:N}^{[2]}|W_1,W_2,\cq)\\
			   =&I(A_{1:N}^{[1]};W_1|\cq)+H(A_{1:N}^{[2]}|W_1,\cq)+H(A_{1:N}^{[1]}, \cdots, A_{1:N}^{[M]}|W_1,W_2,\cq)\notag\\
\label{ind51}   &-H(A_{1:N}^{[1]}, \cdots, A_{1:N}^{[M]}|W_1,\cdots,W_M,\cq)-H(A_{1:N}^{[2]}|W_1,W_2,\cq) \\
			   =&I(A_{1:N}^{[1]};W_1|\cq)+I(A_{1:N}^{[1]},\cdots, A_{1:N}^{[M]};W_3,\cdots,W_M|W_1,W_2,\cq)+H(A_{1:N}^{[2]}|W_1,\cq)\notag\\
			   &-H(A_{1:N}^{[2]}|W_1,W_2,\cq) \\
\label{inductive}\leq& I(A_{1:N}^{[1]};W_1|\cq)+I(A_{1:N}^{[1]},\cdots, A_{1:N}^{[M]};W_3,\cdots,W_M|W_1,W_2,\cq)-H(A_{1:N}^{[2]}|W_1,W_2,\cq) \notag\\
          &+\frac{N}{K}\left(NH(A_1|\cq)-L\right)
\end{align} 
where (\ref{ind1}) follows from the reliability constraint (\ref{reliability}), and (\ref{ind2}) follows from the fact that answer strings are deterministic functions of queries and all messages. (\ref{ind3}) follows from the fact that from $(A_{1:N}^{[1]},\cq)$ the user can reconstruct $W_1$, and similarly for message $W_2$ in (\ref{ind4}). (\ref{ind51}) follows from $H(A_{1:N}^{[1]}, \cdots, A_{1:N}^{[M]}|W_1,\cdots,W_M,\cq)=0$ from the reliability constraint, and (\ref{inductive}) follows from Lemma~\ref{conditioningLemma}.  

For the second term in (\ref{inductive}) we have,
\begin{align}\label{mutualInfo}
I(A_{1:N}^{[1]}&,\cdots, A_{1:N}^{[M]};W_3,\cdots,W_M|W_1,W_2,\cq)\notag\\
                &=H(W_3,\cdots,W_M|W_1,W_2,\cq)-H(W_3,\cdots,W_M|A_{1:N}^{[1]},&\cdots, A_{1:N}^{[M]},W_1,W_2,\cq) \\
                &=H(W_3,\cdots,W_M)\\
\label{second}  &=(M-2)L
\end{align}
Similarly, for the first term in (\ref{inductive}) we have,
\begin{align}\label{third}
I(A_{1:N}^{[1]};W_1|\cq)=L
\end{align}
By using the independence of the first $K$ answer strings, we can lower bound the third term in (\ref{inductive}) using Lemma~\ref{lemma2} as,
\begin{align}\label{IndepW1W2}
H(A_{1:N}^{[2]}|W_1,W_2,\cq) &\geq H(A_{1:K}^{[2]}|W_1,W_2,\cq)=KH(A_1|W_1,W_2,\cq)
\end{align}
Combining (\ref{inductive}), (\ref{second}), (\ref{third}), and (\ref{IndepW1W2}),  we obtain the following upper bound\footnote{A main step of the overall proof is to obtain (\ref{induction_relation}). In the derivation in (\ref{indstart})-(\ref{induction_relation}), we have followed the general spirit of the proof in \cite{JafarPIR} and generalized it to the case of coded databases. We note that this proof can be significantly shortened as shown by the alternative proof in Appendix~\ref{appendixA}.},
\begin{align}\label{induction_relation}
L \leq \frac{N}{K}\left(NH(A_1|\cq)-L\right)-KH(A_1|W_1,W_2,\cq)
\end{align}
which leads to
\begin{align}
\left(1+\frac{N}{K}\right)L &\leq \frac{N}{K}NH(A_1|\cq)-KH(A_1|W_1,W_2,\cq)
\end{align}
Hence, we have the following induction relation,
\begin{equation}\label{GeneralM}
NH(A_1|\cq) \geq \left(1+\frac{K}{N}\right)L+\frac{K^2}{N}H(A_1|W_1,W_2,\cq)
\end{equation}

The relation (\ref{GeneralM}) is the desired induction step as it forms a relationship between the original problem and a reduced PIR problem with $(M-2)$ messages. We note that this relation includes the induction relation in \cite{JafarPIR} as a special case with $K=1$. 

We state the induction hypothesis for $M$ messages as follows, 
\begin{align}\label{induction_hypothesis}
NH(A_1|\cq) \geq L\sum_{i=0}^{M-1} \left(\frac{K}{N}\right)^i
\end{align}

We proved this relation for $M=2$ in (\ref{inductM=2}) as the base induction step. Now, assuming that this is true for $M$ messages, we will prove it for $(M+1)$ messages based on (\ref{GeneralM}) and (\ref{induction_hypothesis}). Since $H(A_1|W_1,W_2,\cq)$ represents $H(A_1|\cq)$ for a reduced PIR problem with $(M-1)$ messages, from the induction hypothesis, we have,
\begin{align}
NH(A_1|W_1,W_2,\cq) \geq L\sum_{i=0}^{M-2} \left(\frac{K}{N}\right)^i
\end{align} 
Substituting this in (\ref{GeneralM}), 
\begin{align}
NH(A_1|\cq) 
        &\geq \left(1+\frac{K}{N}\right)L+L \cdot \frac{K^2}{N^2} \sum_{i=0}^{M-2}\left(\frac{K}{N}\right)^i\\
		&\geq L \sum_{i=0}^M \left(\frac{K}{N}\right)^i
\end{align}  
which concludes the induction argument. Consequently, the upper bound for the coded PIR problem starting from (\ref{PIRrate}) is,
\begin{align}
R&=\frac{L}{\sum_{n=1}^{N} H(A_n^{[1]})}\\
 & \leq \frac{L}{NH(A_1|\cq)} \\
 \label{resultM}&=\frac{1}{\sum_{i=0}^{M-1} \left(\frac{K}{N}\right)^i} \\
 &=\frac{1}{\sum_{i=0}^{M-1} R_c^i}=\frac{1-R_c}{1-R_c^M}
\end{align}
where (\ref{resultM}) follows from (\ref{induction_hypothesis}).
\section{Conclusions}

In this paper, we considered the private information retrieval (PIR) problem over coded and non-colluding databases. We employed information-theoretic arguments to derive the optimal retrieval rate for the desired message for any given $(N,K)$ storage code. We showed that the PIR capacity in this case is given by $C=\frac{1-R_c}{1-R_c^M}$. The optimal retrieval rate is strictly higher than the best-known achievable scheme in the literature for any finite number of messages. This result reduces to the capacity of the classical PIR problem, i.e., with repetition-coded databases, by observing that for repetition coding $R_c=\frac{1}{N}$. Our result shows that the optimal retrieval cost is independent of the explicit structure of the storage code, and the number of databases, but depends only on the code rate $R_c$ and the number of messages $M$. Interestingly, the result implies that there is no gain of joint design of the storage code and the retrieval procedure. The result also establishes a fundamental tradeoff between the code rate and the PIR capacity. 

\appendix
\section{Alternative Proof for (\ref{induction_relation})} \label{appendixA}
\begin{align}
			L&=H(W_2)\\
\label{app1} &=H(W_2|W_1,\cq)\\
\label{app2} &=H(W_2|W_1,\cq)-H(W_2|A_{1:N}^{[2]},W_1,\cq)\\
			 &=I(A_{1:N}^{[2]};W_2|W_1,\cq)\\
			 &=H(A_{1:N}^{[2]}|W_1,\cq)-H(A_{1:N}^{[2]}|W_1,W_2,\cq)\\
\label{app3} &\leq \frac{N}{K}\left(NH(A_1|\cq)-L\right)-H(A_{1:K}^{[2]}|W_1,W_2,\cq)\\
\label{app4} &=\frac{N}{K}\left(NH(A_1|\cq)-L\right)-KH(A_1|W_1,W_2,\cq)
\end{align}
where (\ref{app1}) follows from the independence of $(W_1,W_2,\cq)$, (\ref{app2}) follows from the reliability constraint on $W_2$, (\ref{app3}) follows from Lemma~\ref{conditioningLemma} and the non-negativity of the entropy function, and  (\ref{app4}) follows from the independence of any $K$ answer strings as proved in Lemma~\ref{lemma2}.  
\bibliographystyle{unsrt}
\bibliography{references}
\end{document}